\documentclass{optica-article}

\journal{opticajournal} 

\articletype{Research Article}

\usepackage{lineno}

\usepackage{siunitx}
\usepackage{multicol}
\usepackage{bicaption}
\usepackage[para,online,flushleft]{threeparttable}
\usepackage{soul}
\usepackage{moresize}
\usepackage{lscape}
\linenumbers 

\begin{document}
\nolinenumbers
\title{Asymmetric split-ring plasmonic nanostructures for  optical sensing of \textit{Escherichia coli}} 

\author{Domna G. Kotsifaki,\authormark{1,2,*} Rajiv Ranjan Singh,\authormark{3} S\'{i}le {Nic Chormaic},\authormark{1,*} Viet Giang Truong\authormark{1}}

\address{\authormark{1}Light-Matter Interactions for Quantum Technologies Unit, Okinawa Institute of Science and Technology Graduate University, Onna, 904-0495, Okinawa, Japan\\
\authormark{2}Division of Natural and Applied Sciences, Duke Kunshan University, Kunshan, 215316, Jiangsu Province, China\\
\authormark{3}Information Processing Biology Unit, Okinawa Institute of Science and Technology Graduate University, Onna, 904-0495, Okinawa, Japan}

\email{\authormark{*}dk310@duke.edu;domna.kotsifaki@dukekunshan.edu.cn} 


\begin{abstract*} 
Strategies for in-liquid micro-organism detection are crucial for the clinical and pharmaceutical industries. While Raman spectroscopy is a promising label-free technique for micro-organism detection, it remains challenging due to the weak bacterial Raman signals. In this work, we exploit the unique electromagnetic properties of metamaterials to identify bacterial components in liquid using an array of Fano-resonant metamolecules. This Fano-enhanced Raman scattering (FERS) platform is designed to exhibit a Fano resonance close to the protein amide group fingerprint around 6030~nm. Raman signatures of~\textit{Escherichia coli} were recorded at several locations on the metamaterial under off-resonance laser excitation at 530~nm, where the photodamage effect is minimized. As the sizes of the~\textit{Escherichia coli} are comparable to the micro-gaps~\textit{i.e} 0.41~$\mu$m, of the metamaterials, its local immobilisation leads to an increase in the Raman sensitivity. We also observed that the time-dependent FERS signal related to bacterial amide peaks increased during the bacteria's mid-exponential phase while it decreased during the stationary phase. This work provides a new set of opportunities for developing ultrasensitive FERS platforms suitable for large-scale applications and could be particularly useful for diagnostics and environmental studies at off-resonance excitation.

\end{abstract*}

\section{Introduction}

Bacteria and other micro-organisms are responsible for many human diseases.  Their rapid and accurate identification is crucial for effective treatment and the prevention of further infections. At present, the primary diagnoses of bacterial infection are culture-based methods such as polymerase chain reaction (PCR)~\cite{Belgrader} and enzyme-linked immunosorbent assay (ELISA)~\cite{Dylla}. While culturing methods produce highly accurate results, such approaches are time-consuming, and require complex pretreatment of the sample. 
Hence, there is a strong need for alternative methods displaying more convenient, rapid, and sensitive features for bacterial detection and identification. Raman spectroscopy has proven to be a useful tool to identify bacterial species~\cite{Ute,Stephan}. However, high concentrations of bacteria are required for accurate assignment of several relevant weak vibration bands~\cite{Ahmad}. 
Recently, a machine learning algorithm has been developed to mitigate the low signal-to-noise ratios from bacterial Raman spectra~\cite{ChiSingHo}. Although this approach has the potential for culture-free pathogen identification, the deep learning algorithms are highly data-demanding leading to low algorithm performance.

Meanwhile, the ability to manipulate and control light below the diffraction limit using patterns based on various geometries of metallic elements~\cite{KotsifakiChormaic2019} has already been utilised for biosensing~\cite{Kabashin, Salazar, benesova, wang} and particle trapping~\cite{Sergides_2016, Gordon}. As widely reported in the literature, the main characteristics of these sensing devices, such as sensitivity, the limit of detection (LOD), and repeatability, depend on their plasmonic properties~\cite{Yang}. These properties can be controlled and tuned by modifying the shape or size of a unit cell of the pattern. Among the various optical biosensor schemes, metamaterial-based plasmonic biosensors~\cite{Ahmad,Kabashin,Dutta} show great potential because of their exotic effects, such as negative refraction~\cite{Smith}, perfect lensing~\cite{Zheludev}, and even optical cloaking~\cite{Pendry}, enabling unprecedented control of light at the subwavelength scale. In addition, metamaterials operating at THz frequencies~\cite{Ahmad} have micron-sized apertures and can serve as ideal platforms for fungal and bacterial detection since the sizes of these micro-organisms are on the order of~$\lambda$/100~($\lambda$ being the wavelength of laser), which is comparable to the micro-aperture size. 

Similarly, Fano-resonant asymmetric metamaterials~\cite{Boris} have been utilised for biosensing~\cite{Wu} and  nanoparticle trapping~\cite{Kotsifaki1,DK2, Bouloumis_Nano}. Fano-like resonance structures possess an inherent sensitivity to changes in the local environment owing to the strong interference between super-radiant and subradiant plasmonic modes~\cite{Boris}. 
In addition, periodically arranged metamolecules consisting of nano-aperture structures are characterized by strong localized electromagnetic fields~\cite{Kotsifaki1}, enabling the detection of small amounts, as low as fg/mL, of chemical and biological substances~\cite{Wu}. 
The diffractive scattering of photons by each metamolecule can excite localized surface plasmons (LSP) of adjacent metamolecules rather than decaying into free space. This leads to an additional suppression of the radiative loss by the lattice resonances~\cite{Baptiste, Odom}. Consequently, such sensitive properties make Fano-like structures particularly attractive as biomedical and chemical sensing platforms compared with traditional sensing schemes such as culture-based methods~\cite{Baptiste,Agusti} due to the fact that they are non-labeling, non-destructive, and faster. 

Aside from the above, surface-enhanced Raman spectroscopy (SERS) offers a potential solution for label-free cellular identification with minimal sample preparation and low cellular damage. Despite considerable advances in the characterization of bacteria using SERS, many previous studies used dried samples and long spectral acquisition times of a few minutes. Clinical and pharmaceutical samples are generally in liquid and drying can remove or modify important information related to the SERS spectra. In addition, the majority of SERS-based bacterial biosensing has been performed with plasmonic nanoparticles~\cite{Tadesse,Zhou}, where nonuniform nanoparticle distribution can lead to spot-to-spot variations in the SERS enhancement, making quantification difficult. The generation of reproducible SERS spectra remains a challenging task for some complex biospecimens, such as bacteria. Even though advances with SERS are promising, new approaches should improve accuracy, reproducibility, and effectiveness to be suitable for pharmaceutical and clinical control. A solution may be found by using nanostructure arrays~\cite{Kotsifaki1,DK2} with long-range ordered features that eliminate spot-to-spot spatial variation, thence leading to homogeneous enhanced Raman scattering. 

Here, we demonstrate a liquid bacterial Fano-Resonant Enhanced Raman Spectroscopy (FERS) platform with large-area FERS enhancement and sensitivity for the amide modes of proteins. We designed and fabricated an array of asymmetric split-rings (ASRs) consisting of two types of nano-apertures,~\textit{C}-type and~\textit{l}-type, that support a Fano-resonant mode, allowing for strong background suppression, increased sensitivity, and significant field enhancement. Fano-resonant biosensors can provide accurate detection of low-weight biological specimens at fairly low concentrations. Our metamaterial is designed to exhibit a Fano resonance peak at 6030~nm, which is spectrally in the proximity of the amide vibrational modes. The spectra were experimentally measured at 1$\times$10$^{4}$ CFU/ml of~\textit{Escherichia coli} concentration for the mid-exponential phase (MEP) while a concentration of 1$\times$10$^{8}$ CFU/ml was used for the stationary phase (SP). Although the excitation laser at 532~nm was far from the resonant peak, the Raman signature of bacterial contents was clearly observed. A spectral variation over time was also observed, with an increase of amide peak intensity during the MEP and a decrease during the SP. This metamaterial approach achieves sufficient detection of the amide peaks by using off-resonance excitation, paving the way for the realization of ultrasensitive bacterial identification techniques. 

\section{Methods}
\begin{figure}[ht]
\centering
\includegraphics[trim={0cm 0cm 0cm 0.3cm},clip, width=1\textwidth]{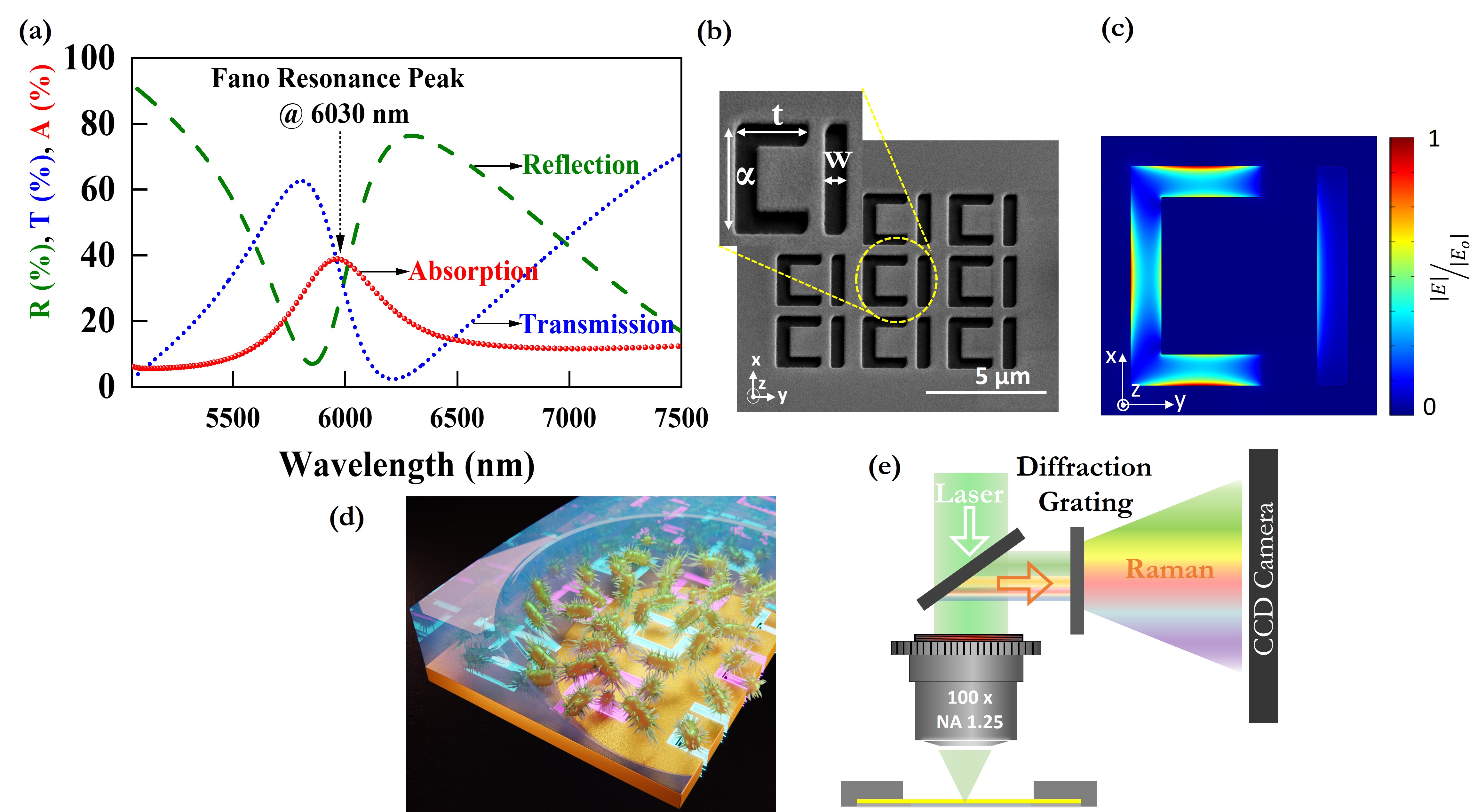}
\setlength\abovecaptionskip{0pt}
\caption{\label{Fig.1} (a) Theoretical reflection, transmission, and absorption spectra for the metamaterial device, showing the resonance peak at 6030~nm close to amide protein vibrations (1676~cm$^{-1}$). (b) Scanning electron microscope (SEM) image, viewed at $52^{\circ}$ from the surface normal of the metamaterial. The geometrical dimensions of each metamolecule unit (enlarged image) are: vertical slit \textit{$\alpha$} = 2.8\,~$\pm\,~0.3$~$\mu$m, horizontal slit \textit{t} = 1.7\,~$\pm\,~0.2$~$\mu$m, slit width \textit{w} = 0.41\,~$\pm\,~0.02$~$\mu$m, and periodicity~\textit{p}~=~3.6\,~$\pm\,~0.2$~$\mu$m. (c) The electromagnetic field enhancement of a single metamolecule at a simulated resonance of 6030~nm for the \textit{y}-direction. The plasmonic hotspots located at the edges of the \textit{C}-type nano-aperture. (d) An illustration image of ~\textit{E.coli} on the metamaterial. (e) Schematic illustration of the experimental setup to collect the Raman spectrum of bacteria immobilized on the metamaterial.}  
\end{figure}

As a first step, to determine the theoretical absorption spectral peak of the metamaterial used in this study, we applied the finite element method using the COMSOL Multiphysics software package~\cite{EPA}. The spatial metamolecule geometric characteristics from our simulations are noted in the caption of Figure~\ref{Fig.1}. Periodic boundary conditions were imposed in the $x$- and $y$-directions to account for the periodic arrangement of the metamolecules, and the array was modeled using Floquet periodicity. Absorption peak spectra are used as indicators of the theoretical resonance position of the metamaterial device. We designed our metamaterial to operate in the THz regime because biological samples possess mid-infrared vibrational fingerprints that can be used for their identification, leading to improved biodetection specificity. In addition, the micron-sized gap of our metamaterial is compatible with the size of an~\textit{E.coli} (radius:~0.5~$\mu$m), allowing for local bacterium immobilization and thus an increase in sensitivity. The Fano resonant peak is noted in Figure~\ref{Fig.1}(a) at 6030~nm, which is in close proximity to protein amide group vibrations~\cite{Wu,Zhou}. Except for the Fano resonance, we did not observe any cavity modes between 500~nm and 7500~nm. 

Our Raman spectroscopy system (3D Laser Raman Microspectrometer Nanofinder 30) consists of a Nd:YAG laser beam ($\lambda$ = 532~nm with maximum incident power of 17~mW) focused using a high numerical aperture (NA = 1.25) oil immersion objective lens (Plan-Neofluar 100×, Carl Zeiss) on to the metamaterial. A monochromator with a 1800~grooves/mm grating was used for spectra collection. 
Figure~\ref{Fig.1}(b) shows a scanning electron microscope (SEM) image of the metamaterial device. It consists of an array of 17$\times$17 ASRs and was fabricated using focused ion beam milling on a 50~nm thick gold film. The zoomed in SEM image shows the metamolecule's geometrical characteristics of periodicity,~\textit{p}~=~3.6\,~$\pm\,~0.2$~$\mu$m, and gap size, 0.41\,~$\pm\,~0.3$~$\mu$m (Fig.~\ref{Fig.1}(b)). The metamaterial was sealed with a glass cover slip and an adhesive microscope spacer of 10~$\mu$m to form a microwell. One micro-liter of bacterial solution was pipetted into the well and the device was mounted on top of a piezoelectric translation stage, as shown in Figure~\ref{Fig.1}(c). We used~\textit{E. coli} BL21 cells, without any plasmid construct for antibiotic resistance. The primary culture was set up in Luria-Bertani (LB) broth medium and cultured overnight in a shaking incubator set at 25$^{\circ}$C and 180~rpm. The following day, the overnight culture was subcultured in LB broth without any antibiotics (0.5 to 1\% inoculation) and grown to mid-exponential phase (MEP) (optical density or OD 0.5 to 0.6) in a shaking incubator set at 37$^{\circ}$C and 180~rpm. The \textit{E. coli} cultures were diluted 1/1000 to a final concentration of 10$^{4}$ CFU/ml before FERS observation~\cite{Yanyu}. Note that during the stationary phase (SP) with OD 1.2, a  bacterial concentration of 1$\times$10$^{8}$ CFU/mL was used to collect the FERS signal.~\textit{E. coli} is a Gram-negative bacterium and interacts strongly with gold nanostructures via lipopolysaccharide carboxylate groups~\cite{Evelin}. 

\begin{figure}[ht]
\centering
\includegraphics[trim={0cm 0cm 0cm 0.3cm},clip, width=0.8\textwidth]{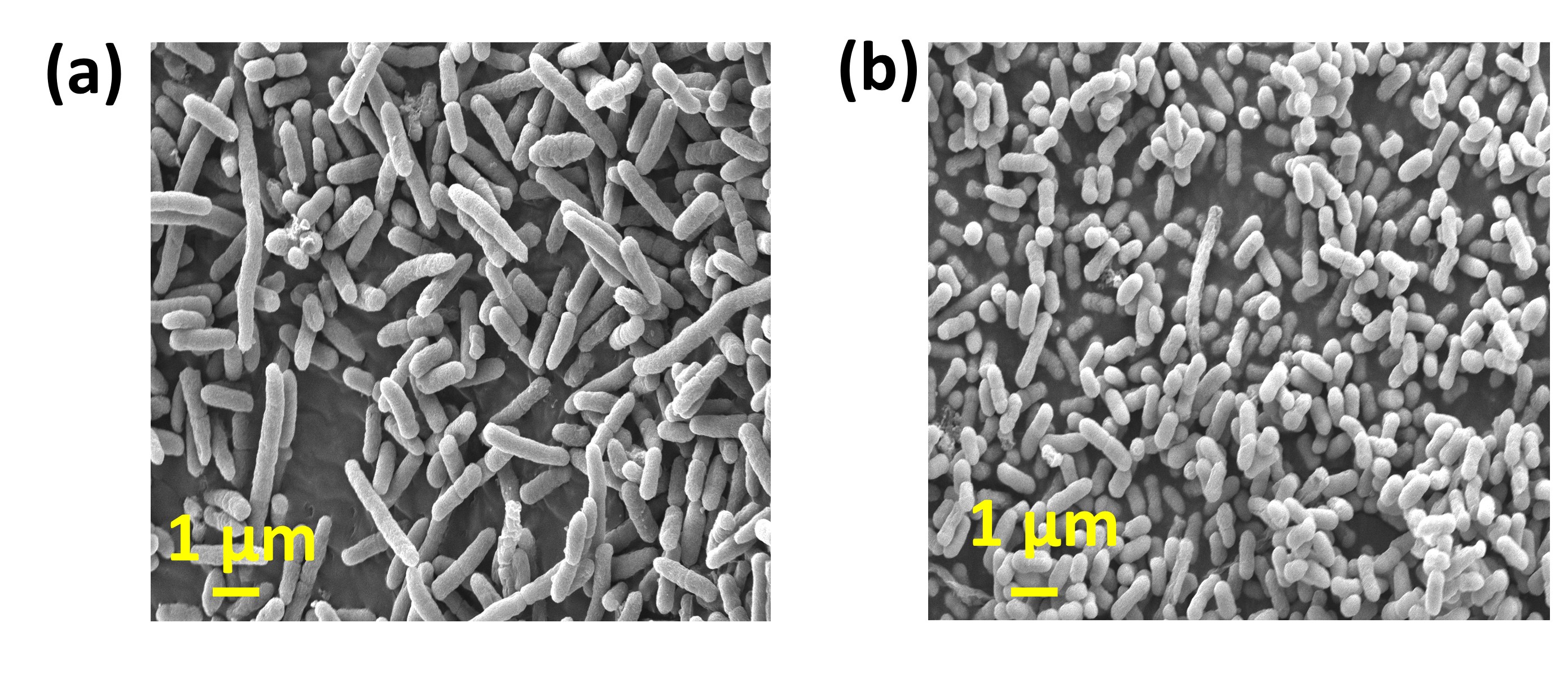}
\setlength\abovecaptionskip{5pt}
\caption{\label{Fig.2} SEM images showing the morphology of ~\textit{E.coli} strains grown to the exponential phase in LB broth cultures at 37$^{\circ}\mathrm{C}$ and 180 rpm shaking in an incubator shaker. (a) Mid-exponential phase and (b) stationary phase of \textit{E. Coli}.
}  
\end{figure}

Figure~\ref{Fig.2} shows scanning electron microscopy (SEM) images of the \textit{E.coli} morphology in (a) mid-exponential and (b) stationary phases.  For sample preparation, wild-type ~\textit{E. coli} BL21 cells were cultured in LB broth at 37$^{\circ}$C for 3.5 to 4~hours for the exponential phase (OD 0.6) and 12~hours for the stationary phase (OD 1.6-1.8). The bacterial suspension was then centrifuged at 5,000~$\times$~g for 10~min. After the supernatant was removed, all  processing was done at 4$^{\circ}$C. The cells were thrice cleaned with 0.40~M HEPPES buffer (pH 7.4), then primary fixation solution (2\% glutaraldehyde) was added, and the cells were fixed for 2 h. After twice washing with HEPES  within 20~min, the cells were fixed with 1\% osmium tetroxide in HEPES buffer  for 2~h, again twice washed with distilled water in 20~min, then stained with 0.5\% Uranyl acetate overnight. Next, the cells were gradually dehydrated with different concentrations of ethanol (20, 30, 40, 60, 70, 80, 90 and 100\%) and then washed in replacement solution (100\% tert-butanol) for 20~min. Finally, the cells were kept in 100\% tert-butanol at -20$^{\circ}$C for 2 h then freeze-dried for 1~h (via Jeol JFD-320 Freeze Drying device). Before imaging, the cells were electrically treated and sprayed with 10~nm osmium tetroxide (via Filgen Osmium Plasma Coater) then imaged using an SEM. In the MEP, bacteria have an elongated rod-shape morphology and most are dividing (see Fig.~\ref{Fig.2}(a)). We observed that bacteria in the SP have relatively short rod shapes, some have spherical shapes and relatively thick periplasmic spaces, and only a few are dividing.  During this phase, several bacteria have rough surfaces and are undergoing death (Fig.~\ref{Fig.2}(b)). Note that the SP is a state of low metabolic activity in which bacteria are protected from starvation and other stresses for long periods of time~\cite{Yanyu}. In this phase, bacteria cease to divide but can still recover when nutrient levels improve~\cite{Yanyu}. 

\section{Results and Discussion}

\begin{figure}[h!]
\centering
\includegraphics[trim={0cm 0cm 0.5cm 0cm},clip, width=0.55\textwidth]{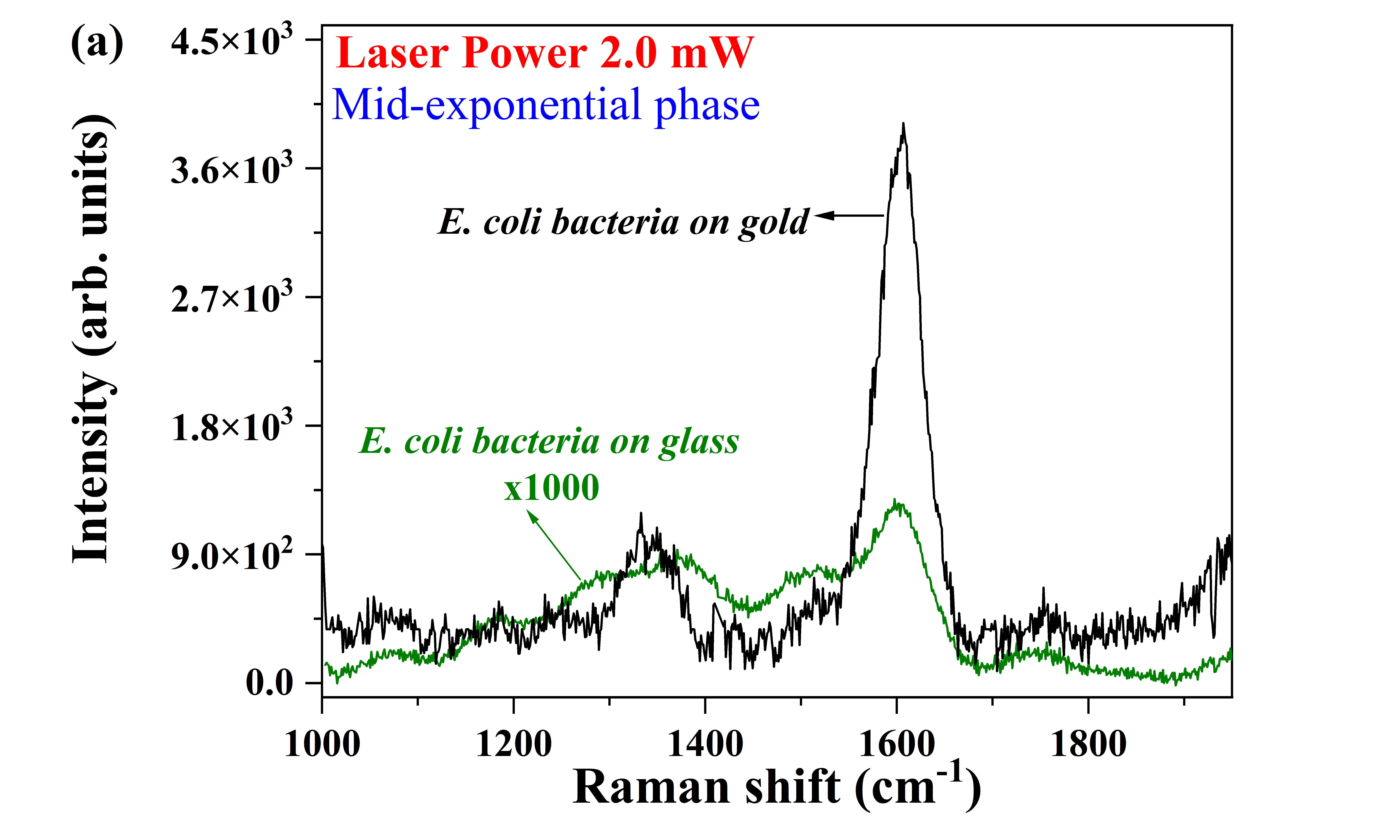}
\includegraphics[trim={0cm 0cm 0cm 0cm},clip, width=0.55\textwidth]{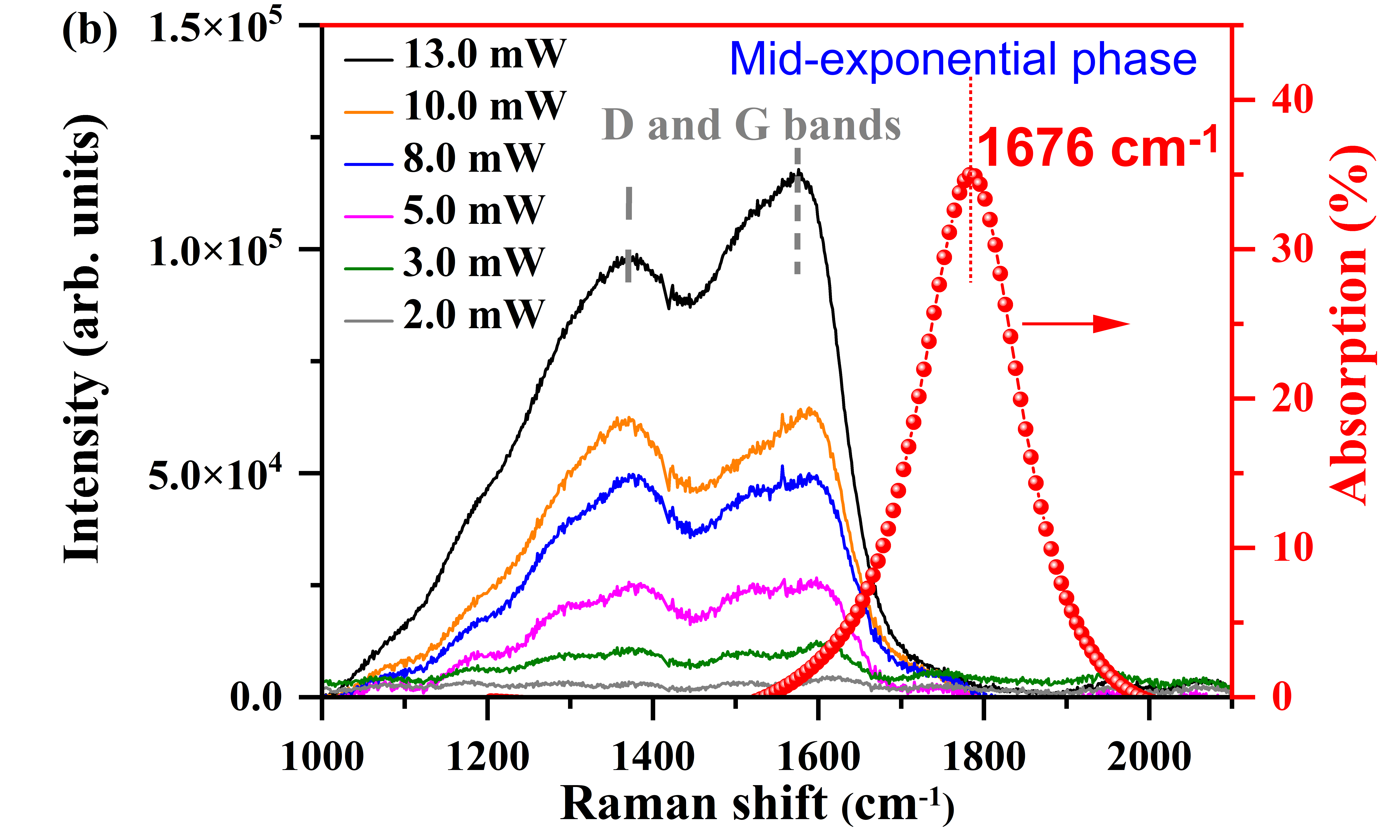}
\includegraphics[trim={0cm 0cm 0.5cm 0cm},clip, width=0.55\textwidth]{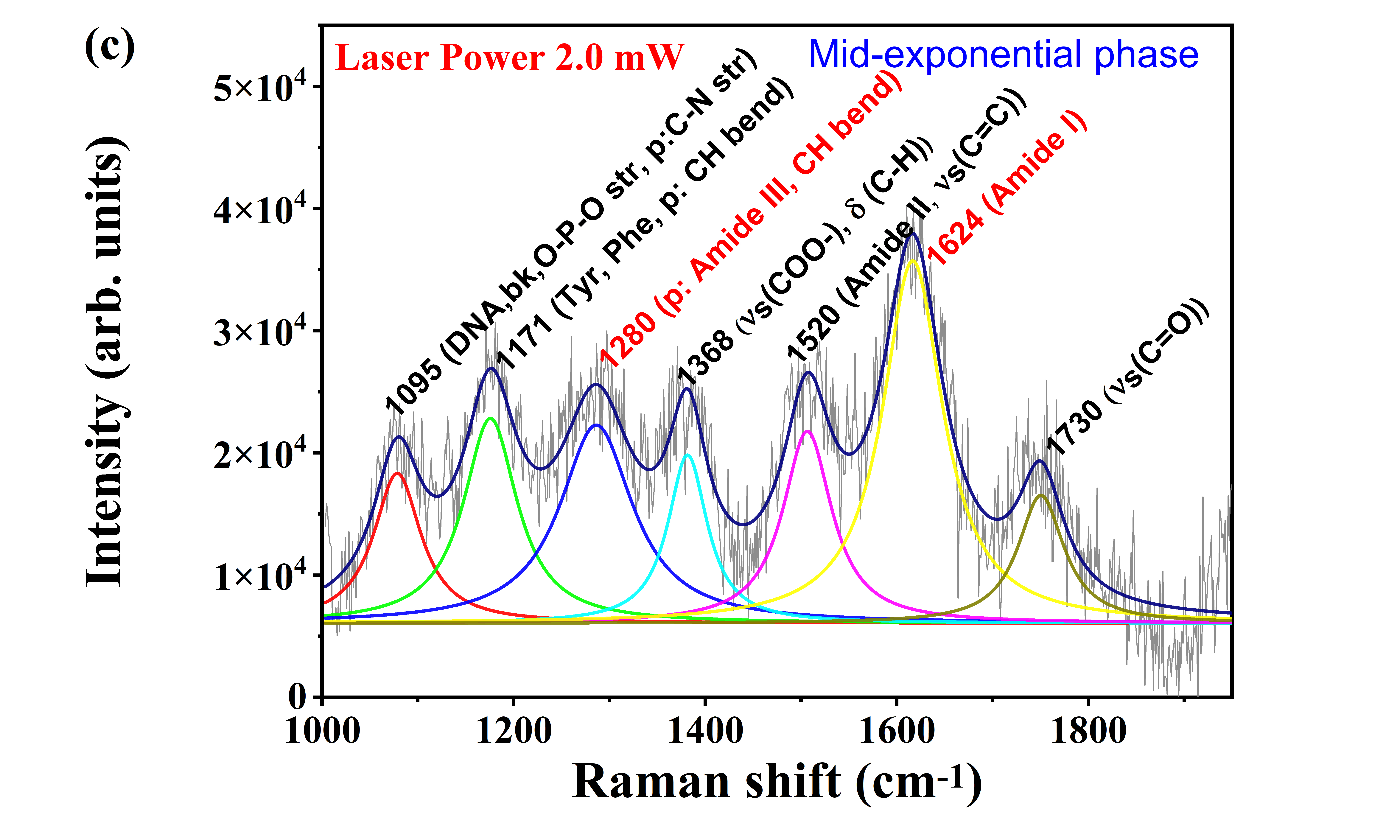}
\setlength\abovecaptionskip{10pt}
\caption{\label{Fig.3} (a) Raman signal of~\textit{E. coli} bacteria in the MEP on gold (black line) and microscope glass (green line $\times$1000) substrates. The Raman signal from the glass substrate is rescaled by a factor of 1000 for ease of viewing. (b) Raman spectra of~\textit{E. coli} bacteria in the MEP on the metamaterial for different laser powers measured before the objective lens.  The D and G bands of amorphous carbon are noted at 1365~cm$^{-1}$ and 1567~cm$^{-1}$, respectively. The theoretical absorption spectrum is also presented in red. (c) Raman spectrum of~\textit{E. coli} bacteria in the MEP on the metamaterial (grey line/Fig.~\ref{Fig.3}(b)) using 2.0~mW incident laser power. The solid colored lines are from Lorentz fitting. The measurements were performed in liquid and under the same experimental conditions (wavelength 532~nm, objective lens 100x, integration time 10~s and accumulation equal to one). The bacterial concentration for the MEP was 10$^{4}$ CFU/mL.}
\end{figure}

In Fig.~\ref{Fig.3}(a) we can see the measured Raman spectrum from the MEP bacteria (OD: 0.5 to 0.6 and diluted bacterial concentration of 10$^{4}$ CFU/mL) on 50~nm gold film and a glass substrate.  We notice that the spectra do not display the same peaks as those obtained with the metamaterial under the same interrogation settings; we hypothesise that the spectral peaks are attributable to enhancement from the Fano-resonant mode supported by the metamaterial. 
Figure~\ref{Fig.3}(b) shows Raman spectra from the MEP of~\textit{E.coli} on the metamaterial recorded for different laser powers measured before the objective lens. At laser powers higher than 3.0~mW, two broad, strong intensity bands located at 1365~cm$^{-1}$ and 1567~cm$^{-1}$ are observed. The photo-induced degradation of biological samples~\cite{Kots} often results in the presence of these characteristic bands in recorded Raman spectra owing to the formation of amorphous carbon~\cite{Evelin}. By using lower incident laser powers, this carbonization effect is minimized, allowing for the Raman signature of~\textit{E. coli} to be recorded.

Figure~\ref{Fig.3}(c) shows a Raman spectrum for ~\textit{E. coli} in the MEP (grey line) on the metamaterial for a laser power of 2.0~mW before the objective lens. Note that the spectrum for each experiment in this work  was collected with 10~s acquisition time while the concentration of bacteria for each measurement was kept constant. We determined the Raman peaks by fitting the experimental spectrum using  Lorentzian functions~\cite{Dholakia}. With laser excitation at 532~nm and a Fano-resonance at around 6030~nm (Fig.~\ref{Fig.1}(a)), signature bacterial amide spectral peaks appear near 1280~cm$^{-1}$ and 1624~cm$^{-1}$. The presence of these bands corresponds with amide III and amide I protein vibrations~\cite{Ivleva,Zhou}. Additionally, Raman peaks arising from components of the bacterial cell membrane, such as phospholipids, liposaccharides, and other polysaccharide moieties, were also observed~\cite{Evelin} (noted in Fig.~\ref{Fig.3}(c)). Furthermore, Raman signals obtained with excitation wavelengths of 532~nm contain high fluorescence backgrounds and cause photodamage to the samples~\cite{Evelin}. However, in this study, the measurements conducted with excitation wavelengths of 532~nm and lower laser incident powers on the metamaterial gave the most satisfactory results, as the obtained FERS spectra showed many bands originating from bacterial components. 
\begin{figure*}
\centering
\includegraphics[trim={0cm 0cm 0cm 0cm},clip, width=1\textwidth]{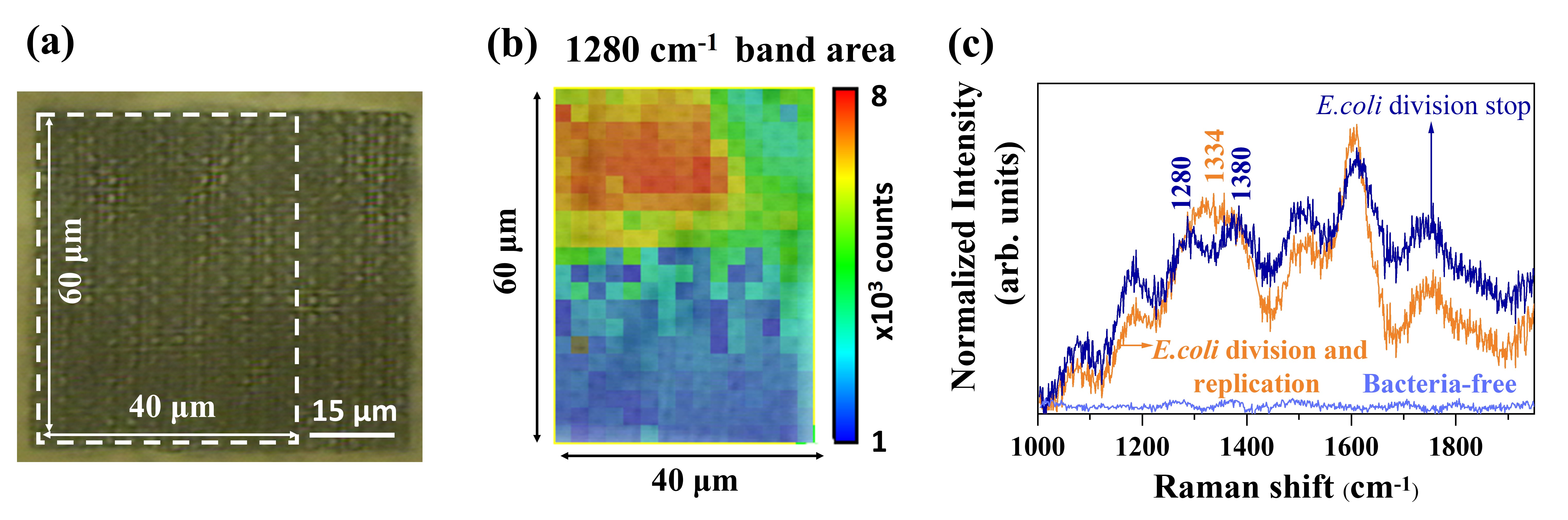}
\setlength\abovecaptionskip{-5pt}
\caption{\label{Fig.4} (a) Microscope image of SP bacterial population on the metamaterial where the distribution of~\textit{E.coli} is noted by the shaded areas. The white dashed rectangle shows the map area. (b) FERS mapping of bacterial population in the SP on the metamaterial in a liquid environment. The red and green regions show metabolically active bacteria whereas blue regions indicate that no bacteria  were etected. (c) FERS spectra after baseline correction and normalization for the red, green, and blue positions noted in (b). The red positions in (b) provide a broad characteristic band at 1334~cm$^{-1}$ that indicates~\textit{CH} deformation vibrations~\cite{Ivleva}. 
The applied concentration of~\textit{E.coli} bacteria was 1$\times$10$^{8}$ CFU/mL (OD:1.2). A 100x objective lens was used and the incident laser power was 3.0~mW.}
\end{figure*}

\begin{figure}[h!]
\centering
\includegraphics[trim={0cm 0cm 0cm 0cm},clip, width=0.65\textwidth]{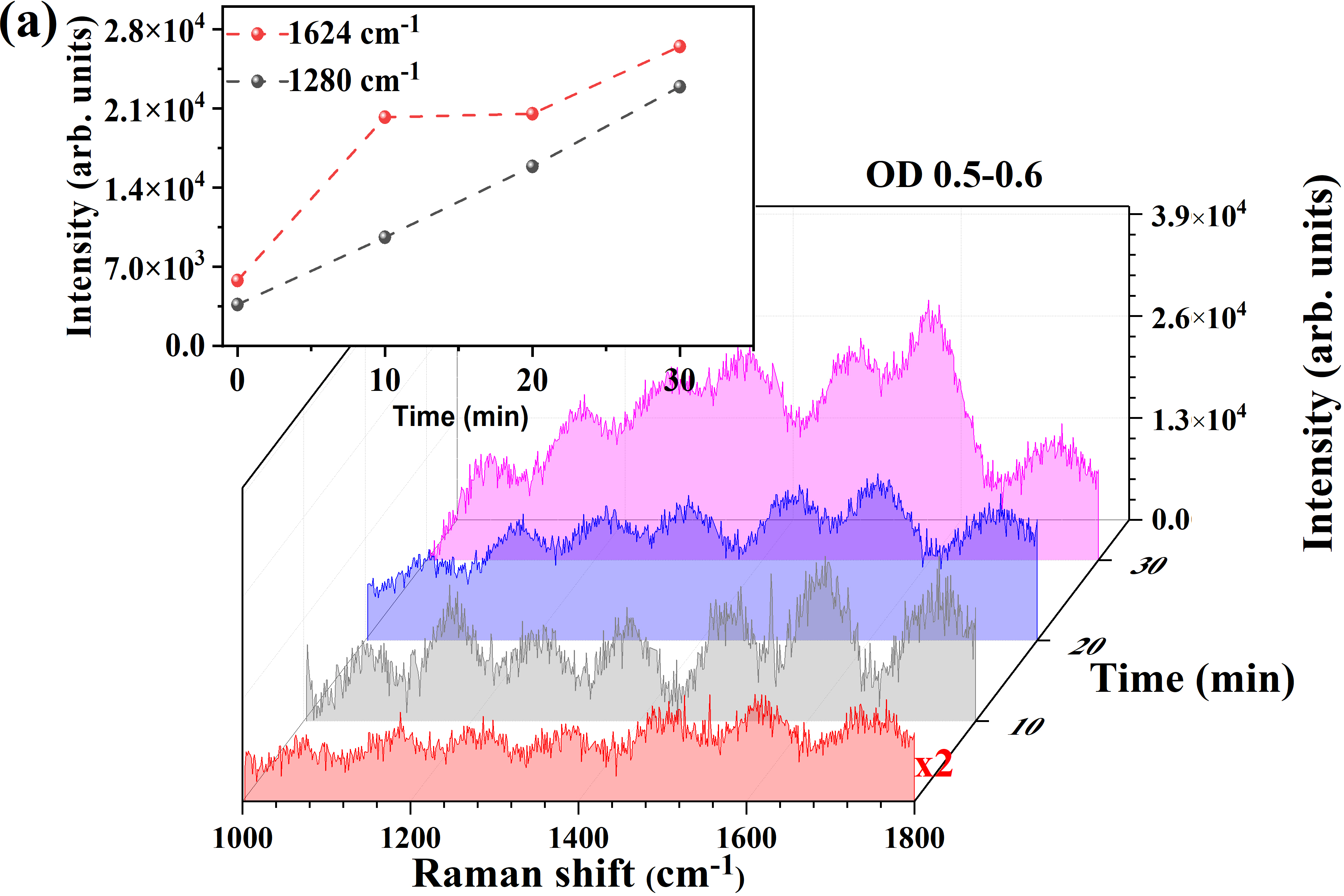}
\includegraphics[trim={0cm 0cm 0cm 0cm},clip, width=0.65\textwidth]{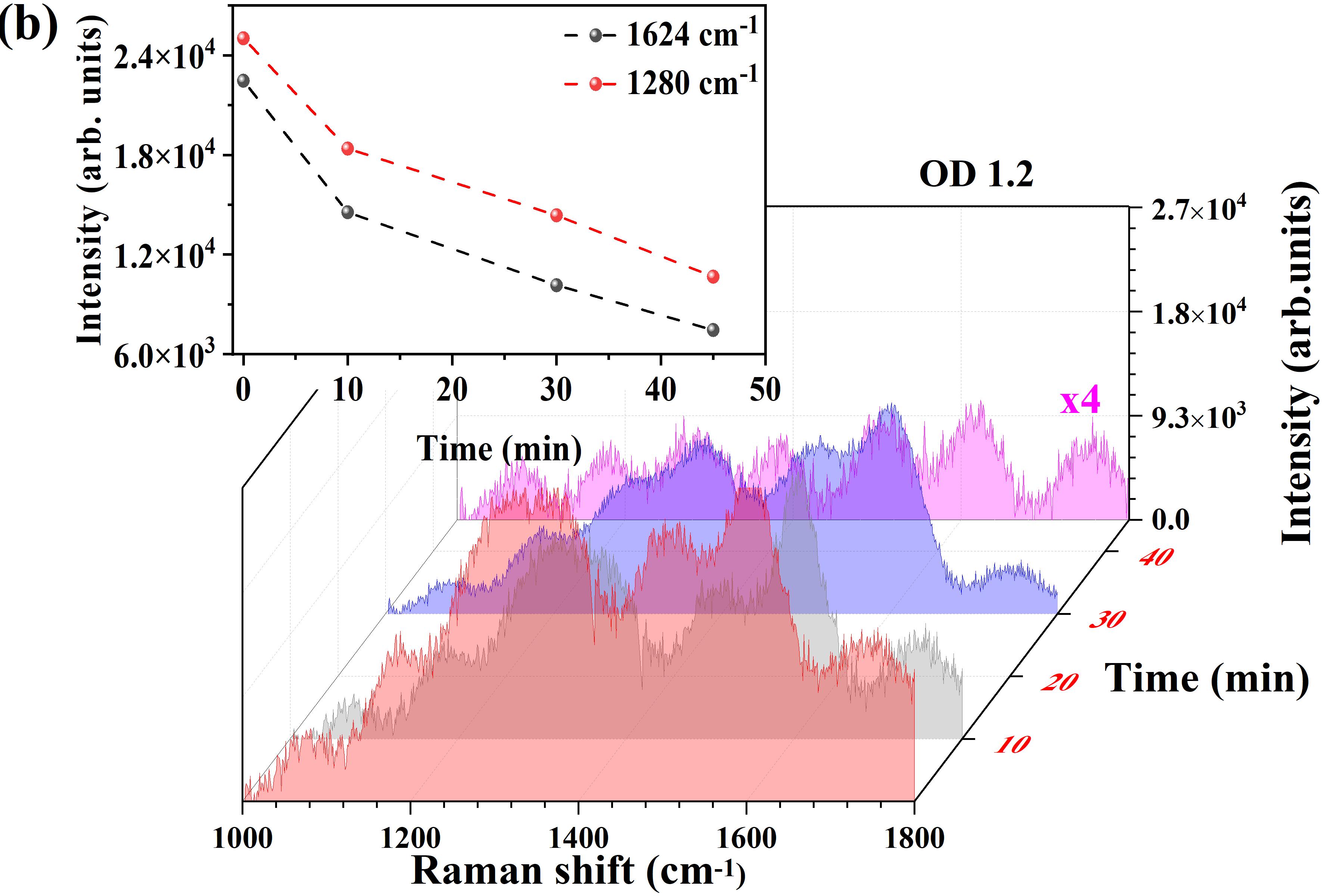}
\includegraphics[trim={0cm 0cm 0cm 0cm},clip, width=0.65\textwidth]{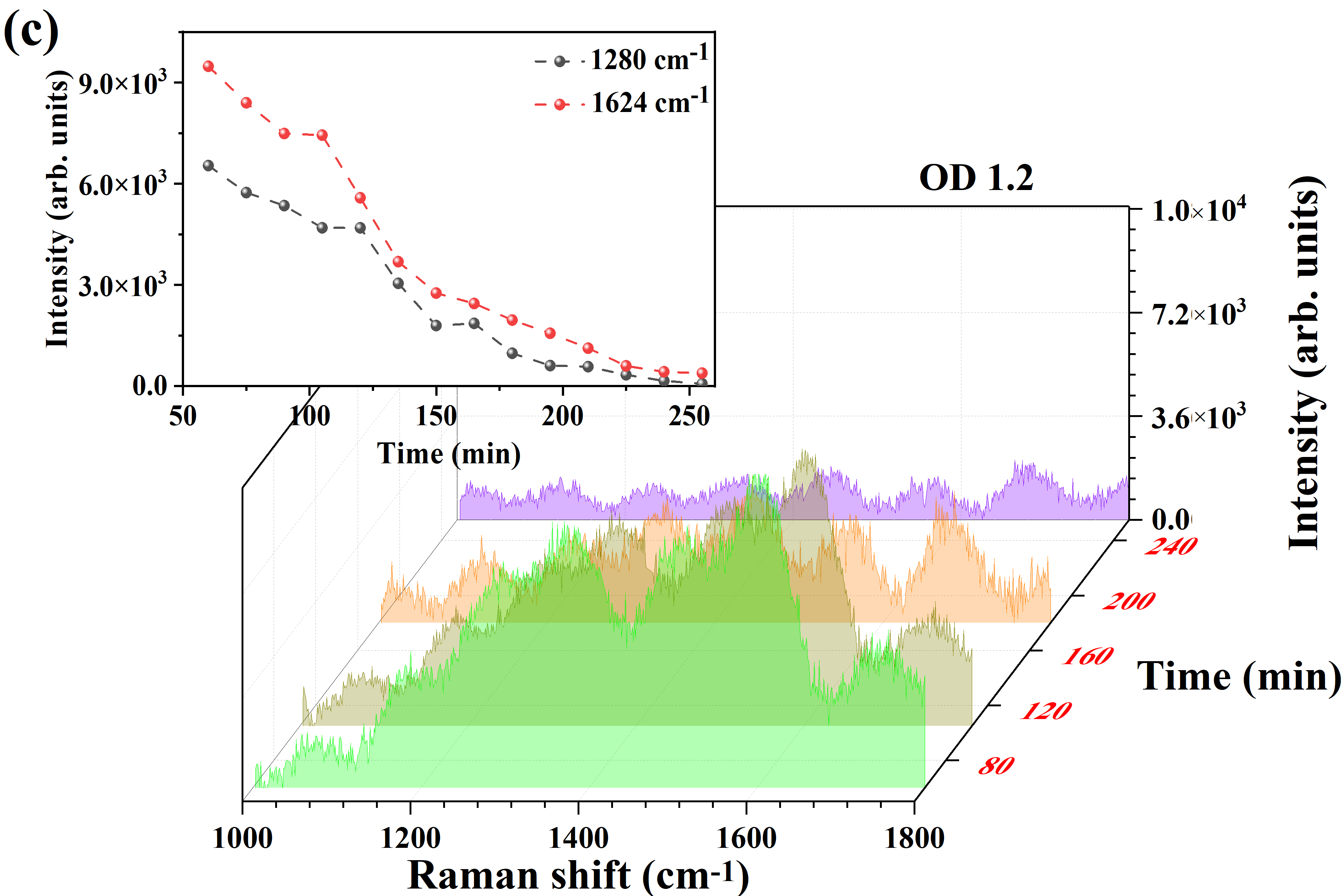}
\setlength\abovecaptionskip{10pt}
\caption{\label{Fig.5} (a) FERS spectra of bacteria in the MEP with an OD: 0.5~-~0.6 on a metamaterial device over time. Inset shows a linear relationship between the amide peak intensity magnitude and time. Spectra from the solution containing  ~\textit{E. coli} in the SP  on a metamaterial device (b) between 0 and 50 min and (c) between 60 and 255 min. Insets: The intensities of the amide characteristic peaks at 1280~cm$^{-1}$ and 1624~cm$^{-1}$ decrease until they approach the values measured on the glass substrate after 255~min. The dashed lines are a guide for the eye.}
\end{figure}

The diagnostic spectral signatures and the intensities of the FERS signals from the bacteria scale with the percentage of metabolically active bacteria in the population. Hence, the FERS experiments were repeated several times for both the MEP and SP of the growth cycle of the bacteria, and at random positions on the metamaterial, to obtain reproducible spectral fingerprints and the corresponding intensity. Since bacteria are living organisms and do not necessarily respond or grow identically over several days, even if the growth conditions are kept the same, we note that the bacterial optical density (OD) of 0.5 to 0.6 or diluted 1$\times$10$^{4}$ CFU/mL used, might be the optimum to provide the most reproducible spectra (Fig.~\ref{Fig.3}(c) and Figs.~\ref{Fig.5}). These observations might have a diagnostic implication for patient care in  clinical settings.

A comparison between the non-FERS (green line in Fig.~\ref{Fig.3}(a)) and FERS (grey line in Fig.~\ref{Fig.3}(c)) spectra of bacteria in the mid-exponential phase revealed that, besides the significant enhancement of the Raman signal, the FERS spectra are characterized by a large number of peaks which is indicative of more chemical information. Although it is very difficult to estimate the enhancement factor for the spectra, from the intensities of the strongest bands and the acquisition conditions used, we assumed the FERS enhancement factor to be equivalent to the magnitude of the ratio between FERS and non-FERS intensities for the same probe and Raman peak. We obtained a FERS enhancement factor on the order of 10$^4$ from a comparison between the intensity magnitude of FERS with the metamaterial and non-FERS values with the glass substrate, while a factor of 10 was noted between the metamaterial and gold substrate at the 1624~cm$^{-1}$ spectral peak. Signal enhancement in SERS can appear when both surface and resonance effects are combined~\cite{Ru}. The resonance effect is high when the plasmon resonance is located between the excitation wavelength and the wavelength that is Raman scattered by the analyte~\cite{Ru, Evelin}. Typical values of SERS enhancement factors (EF)~\cite{Ru} are on the order of 10$^4$-10$^8$. For instance, a plasmonic nanodome array~\cite{WuHsinYu} was fabricated to exhibit an average SERS EF of 8.51$\times$10$^{7}$ while an EF of 4.81$\times$10$^{8}$ was noted using an array of plasmonic nano-mushrooms as the SERS substrate~\cite{XuZhida}. In both cases, the resonance of the structure was located between the excitation wavelength and the Raman scattered wavelength leading to a large enhancement factor.

In our case, the resonance and surface effects were not in close proximity, resulting in a FERS enhancement factor at the lower end of this range. It is well known that two mechanisms~\cite{ikeda}, \textit{i.e.} electromagnetic and chemical effects, can contribute to SERS enhancement. The former, which is primarily responsible for SERS enhancement, occurs when surface plasmons are excited on metallic nanostructures, while charge transfer resonances are caused between a metal state at the Fermi level and a molecular electronic state~\cite{ikeda}. Our signal enhancement observation may therefore be explained by chemical effects. In any case, this is not within the scope of this study.  As discussed above, when using metallic structures with resonances matched to the excitation laser, both the laser and the Raman scattering light are absorbed by the structures themselves. The absorbed photon energy leads to  excitation of the electrons in the metal and the subsequent nonradiative decay of  excited electrons converts the energy to heat. The thermal disturbance caused by plasmonic heating may affect the photochemical degradation of the bacterial cells showing low surface enhanced Raman scattering spectrum reproducibility. In this work, the resonance of the metamaterial is red-shifted from the laser excitation, thereby minimizing the laser-induced heating that can contribute to photodamage of biological entities. 

It is important to note that the Raman cross-section can only be enhanced for molecular components that are sufficiently close (within~10~nm) to the SERS active surface because the electromagnetic enhancement scales with the 12$^{th}$ power of the distance (\textit{d}) between the analyte and SERS substrate~\cite{Kneipp}. Likewise, as the gap size of the metamaterial is comparable to the size of a single bacterium, the latter may be captured in the micro-aperture and could interact strongly with the electromagnetic field, leading to an increase in device sensitivity to the spectral bacterial fingerprints. Therefore, as the penetration depth of the electromagnetic field of the metamaterial is 100 nm, the FERS signal may be derived from the spectral variation of the wall components or/and inner organelles of the bacterium~\cite{Andrei}.

To validate the sensitivity of the metamaterial and collect information on the distribution of bacteria on the surface, FERS mapping was performed using a solution containing bacteria in the stationary phase (OD: 1.2). Generally, during different growth phases, a bacterium produces different biomolecules that reflect its ability to increase cellular constituents from the environment~\cite{Jananee,Neugebauer}. As a result, several Raman fingerprints are generated that reflect different metabolic states. In fact, during  bacterial growth, new proteins are synthesized, whereas when the bacteria division stops the Raman peak intensity assigned to amino acids and proteins decreases~\cite{Jananee}.
A drop of 1~$\mu$L of the SP bacterial solution was placed onto the metamaterial  Figure~\ref{Fig.4}(a) shows a microscope image in which the distribution of SP bacteria on the surface of the metamaterial can be observed (shaded areas).  FERS mapping was performed with a 4~$\mu$m step size, covering an area of 60~$\times$~40~$\mu$m$^{2}$ as shown in Figure~\ref{Fig.4}(b).  To analyze the distribution of various substances in the SP bacterial solution, it was important to choose characteristic frequency regions or marker bands for them.  We chose the SERS band at 1280~cm$^{-1}$ to image the contribution of the amide modes. If no Raman signals were detected, we inferred that the area contained no bacteria. Conversely, if strong Raman signals were obtained, we assumed that the area contained metabolically active bacteria at the time of signal collection. In Figure~\ref{Fig.4}(b) and (c), each red and green position represents the detection of a FERS signal for metabolically active bacteria, whereas the blue positions indicate no bacterial signals were obtained. Moreover, spectral differences between the red and green positions may be explained by the chemical enhancement effect, induced by direct interactions between the analyte and metamaterial. This resulted in a significantly strong vibration near 1334~cm$^{-1}$ (orange spectrum in Fig.~\ref{Fig.4}(c)), which is often assigned to the deformation of~\textit{CH}~mode~\cite{Ivleva} and results in the peak splitting into two: one at 1280~cm$^{-1}$ that is characteristic of amide III vibrations~\cite{Ivleva} and the other at 1380~cm$^{-1}$ that is predominantly due to the symmetric carbohylate stretching mode of polyanionic polysaccharides~\cite{Ivleva}. Note that the peak at 1334~cm$^{-1}$ is often used as a marker of adenine and DNA in SERS analyses of biological specimens~\cite{Efrima}. Based on the above analysis, we assume that the red area in Figure~\ref{Fig.4}(b) represents bacteria that are are dividing and the green area represents bacteria for which the cell division process has slowed down.

Furthermore, we investigated the FERS signals at different times (Fig.~\ref{Fig.5}) and found that the spectra remained largely unchanged throughout the experiment, confirming that the matamaterial can reproduce~\textit{E.coli} Raman spectra. Figure~\ref{Fig.5}(a) shows the time-dependent Raman spectra of~\textit{E.coli} during mid-exponential growth (OD: 0.5~-~0.6) over a 30 min period. We observe that the intensity magnitudes of both characteristic peaks at 1280~cm$^{-1}$ and 1624~cm$^{-1}$ increase with time between 0 and 30~min, demonstrating production and accumulation of bacteria during the MEP. For the investigated sample of MEP bacteria, the 30~min time frame produced the most consistent spectra. Results have been reported in the literature where SERS spectra, obtained from bacterial growth at the start of the exponential phase, heading toward the stationary phase, showed  changes~\cite{Liu2009} in characteristics. Specifically, the intensity of peptide (amino acids and proteins) peaks progressively increased as the bacteria moved from the exponential phase to the MEP~\cite{Liu2009}. In this case, \textit{i.e.} at MEP, the peptidoglycan layer~\textit{i.e.} of the bacterium wall is around 12~nm~\cite{Li} and as the electromagnetic field penetration depth is 100~nm, the intensity of the Raman peaks of the subcellular components increases. However, the thickness of the peptidoglycan layer of the bacterium wall in the stationary phase is larger compared to that during the MEP (around 15~nm~\cite{Li}) and this may lead to a decrease in the Raman peak intensities. 

During the SP, the aspect ratio of rod-like bacteria such as~\textit{E. coli} decreases (Figs.~\ref{Fig.2})~\cite{Yanyu} revealing an additional SERS change~\cite{Liu2009}.
We collected  Raman spectra for bacteria in the SP over 240~min. Compared to bacteria in the MEP, a trend of lower signal intensities toward the end of a long measurement series was observed, see Fig.~\ref{Fig.5} (b) and (c), for bacteria with an OD of 1.2. General, the MEP is characterized by an increase in the growth rate until a constant value is reached~\cite{Huang, Neugebauer}. Once the density of the bacterial population reaches a certain level, the SP begins.  This is where the bacteria start to die and the biological mass no longer increases~\cite{Huang, Neugebauer}. Hence, our observation could be explained by the fact that, as bacteria do not grow during the SP, they may lose the integrity of their cell walls and membranes, which may contribute to recorded FERS spectra, resulting in a decrease in signal intensity. It has been reported that bacteria are unable to divide after 240~min~\cite{Zhou}. 

Note that changing the growth phase may cause differences in spectral properties for the type of ~\textit{E. coli} that we studied, but may not be detectable over time for all bacterial species. This, however, lies outside the scope of the current study. Therefore, the experimental data are consistent with the fact that  bacteria of OD of 0.5~-~0.6 are dividing and are metabolically active during the first 30~min, while their metabolic activity is likely reduced at an OD of 1.2. This may lead to the observed differences in the intensity of the amine peaks. Hence, the time-dependent FERS signals may be directly related to bacterial biochemical changes during different phases. 

\begin{landscape}
\begin{table*}[ht]
\centering
\caption{\textbf{Comparison of this method and other methods based on Raman spectroscopy for detection of ~\textit{E.coli}}}
\label{tab:Table1}
\small
\begin{tabular}{ccccccc}
\hline \hline
\textbf{Sensing Mechanisms} & \multicolumn{1}{l}{\textbf{\textit{Escherichia coli} strain}} &  
\textbf{\begin{tabular}[c]{@{}c@{}}Excitation laser\\ \end{tabular}} &
\textbf{\begin{tabular}[c]{@{}c@{}}Concentration (CFU/mL)\\ \end{tabular}} &
\textbf{\begin{tabular}[c]{@{}c@{}} On-/Off-Resonance\\ \end{tabular}} &
\textbf{\begin{tabular}[c]{@{}c@{}} Enhancement factor\\ \end{tabular}} &
\textbf{\begin{tabular}[c]{@{}c@{}}Reference\\  \end{tabular}} \\
\hline \hline
\begin{tabular}[c]{@{}c@{}} SERS (Ag nanoparticles) \\ [6pt] \end{tabular}  & DSM1116  & 633~nm & $\times 10^{7}$ & on-resonance  & 1.1$\times10^{6}$ $@$ 735 cm$^{-1}$ &~\cite{Zhou}\\
\begin{tabular}[c]{@{}c@{}} SERS (Ag nanosculptured thin films) \\ [6pt] \end{tabular}  & RFM443  & 785~nm & 4.5$\times10^{4}$  & on-resonant & 3.5$\times 10^{3}$ $@$ 1077 cm$^{-1}$ & ~\cite{Srivastava}\\
\begin{tabular}[c]{@{}c@{}} FERS (Au metamaterial-this work) \\ [6pt] \end{tabular}  &  BL21 (B-Strain) & 532~nm & 1$\times 10^{4}$ & off-resonant   & $10^{4}$ $@$ 1658 cm$^{-1}$ (6030~nm) \\
\hline \hline
\end{tabular}
\end{table*}
\end{landscape}

As shown in Table~\ref{tab:Table1}, the metamaterial used in this work may be more promising for identifying bacteria far from resonance compared to SERS. Overall, this study provides promising findings for a specific bacteria,  but the results could differ when different bacterial species are studied, which will be a future study.

\section{Conclusion}

Rapid and reliable identification of pathogenic bacteria is vital in many fields such as health care, food, and environmental sciences. Here, we have demonstrated a FERS platform that can be employed to produce valuable and repeatable bacterial spectral information in a liquid environment. A Fano-resonant ASRs metamaterial was fabricated on a thin 50~nm gold film to identify signature peaks of~\textit{E. coli}. By suitably engineering the Fano lineshape, we produced an efficient FERS-active substrate with spatially localized hotspots and capable of significant FERS enhancement. This enhancement stems from the remarkable sensitivity of metamaterials to minor variations in the dielectric permittivity of the surrounding medium. The influence of the power of the irradiating laser on the Raman signals was investigated to determine the spectral resolution and specificity of the device. Compared to other analytical methods or SERS schemes, our approach opens a new set of opportunities for the development of better-performance FERS substrates that could enable the precise and rapid identification of bacteria, fungi or viruses in liquid. Until now, high bacterial concentrations were used for SERS analysis. In this work, the number of bacteria used in combination with the micro-sized aperture of the metamaterial enables  bacteria immobilization on the substrate. Thus, the device is well-suited for the screening and study of single bacterium. The  metamaterial substrates may place the bacteria into the micron-sized apertures and could be used to study metabolic degradation pathways~\cite{Premasiri} or quorum-sensing in bacteria~\cite{Bodelon}. It is important to mention that since different bacterial species act very differently, the conclusions that are entirely valid in this study may not apply to different bacterial species. We envision that Fano-resonant nanostructures have strong potential to be employed in practical on-chip devices, enabling high specificity detection of biological substances in various environments.  

\backmatter

\section{Acknowledgments}
The authors thank M. Ozer for technical assistance, P. Puchenkov from the Scientific Computing and Data Analysis Section at OIST, and Toshiaki Mochizuki from the Imaging Section at OIST. 

\section{Data Availability}
The data that support the findings of this study are available from the authors upon reasonable request.

\section{Funding}
This work was supported by funding from  Okinawa Institute of Science and Technology Graduate University. DGK acknowledges support from JSPS Grant-in-Aid for Scientific Research (C) Grant Number GD1675001.

\bibliography{sample}

\end{document}